\documentstyle[12pt]{article}
\newlength{\extraspace}
\newlength{\extraspaces}
\newcommand{\be}{\begin{equation}
\addtolength{\abovedisplayskip}{\extraspaces}
\addtolength{\belowdisplayskip}{\extraspaces}
\addtolength{\abovedisplayshortskip}{\extraspace}
\addtolength{\belowdisplayshortskip}{\extraspace}}
\newcommand{\ee}{\end{equation}}
\newcommand{\ba}{\begin{eqnarray}
\addtolength{\abovedisplayskip}{\extraspaces}
\addtolength{\belowdisplayskip}{\extraspaces}
\addtolength{\abovedisplayshortskip}{\extraspace}
\addtolength{\belowdisplayshortskip}{\extraspace}}
\newcommand{\ea}{\end{eqnarray}}
\newcommand{\newsection}[1]{
\vspace{7mm}
\pagebreak[3]
\addtocounter{section}{1}
\setcounter{equation}{0}
\setcounter{subsection}{0}
\setcounter{footnote}{0}
\begin{center}
{\large {\bf \thesection. #1}}
\end{center}
\nopagebreak
\medskip
\nopagebreak
\hspace{3mm}}
\newcommand{\nonu}{\nonumber \\[.5mm]}

\newcommand{\ebase}{e^{\!\!\!\!{\scriptstyle\rightharpoonup}}}
\newcommand{\dbase}{\partial^{\!\!\!{}^{\!\!{\scriptstyle\rightharpoonup}}}}
\newcommand{\Gammao}{{\mit\Gamma}^{\!\!\!{}^{\!{}^{\scriptstyle\circ}}}}
\newcommand{\thetao}{\theta^{{}^{\!\!\!{}^{\scriptstyle\circ}}}}
\newcommand{\Ro}{R^{\!\!\!{}^{\!{}^{\scriptstyle\circ}}}}

\newcommand{\RT}{\widetilde R}
\newcommand{\Go}{G^{\!\!\!{}^{\!{}^{\scriptstyle\circ}}}}

\newcommand{\nablao}{\nabla^{\!\!\!\!\!{}^{{}^{\scriptstyle\circ}}}}
\newcommand{\Do}{{\cal D}^{\!\!\!{}^{\!\!{}^{\scriptstyle\circ}}}}
\newcommand{\Boxo}{\Box^{\!{}^{\!\!\!\!{}^{\scriptstyle\circ}}}}

\begin{document}
\addtolength{\baselineskip}{.7mm}
\thispagestyle{empty}
\begin{flushright}
STUPP--96--144 \\
{\tt gr-qc/9603006} \\
March 1996
\end{flushright}
\vspace*{1cm}
\begin{center}
{\large{\bf Dynamical torsion and torsion potential }} \\[20mm]
{\sc Hong-jun Xie} \\[7mm]
and \\[7mm]
{\sc Takeshi Shirafuji} \\[12mm]
{\it Physics Department, Faculty of Science \\[2mm]
Saitama University \\[2mm]
Urawa, Saitama 338, Japan } \\[20mm]
{\bf Abstract}\\[1cm]
{\parbox{13cm}{\hspace{5mm}
We introduce a generalized tetrad which plays the role of a potential 
for torsion and makes torsion dynamic. 
Starting from the Einstein-Cartan action with torsion, 
we get two field equations, 
the Einstein equation and the torsion field equation 
by using the metric tensor and the torsion potential as independent variables; 
in the former equation the torsion potential plays the role of a matter field. 
We also discuss properties of local linear transformations 
of the torsion potential and give a simple example 
in which the torsion potential is described by a scalar field.}}
\end{center}
\vfill
\newpage
\setcounter{section}{0}
\setcounter{equation}{0}
%
%
\newsection{Introduction}
The gauge theory of gravitational field 
was first proposed by Utiyama [1] 
and later developed by Kibble [2] and Sciama [3]. 
The Einstein-Cartan-Sciama-Kibble (ECSK) theory of gravity 
uses the Einstein-Cartan Lagrangian $\sqrt{-g} R$, 
and it does not require torsion to vanish. 
Rather, the torsion is treated as an independent variable 
along with the metric. The torsion is not dynamic in the ECSK theory, 
however, being algebraically determined by the local spin density: 
Namely, the torsion is always a pointwise function of the source field, 
and it is usually called {\it frozen torsion}.
\par
If we are to consider ``dynamical torsion'' --- torsion 
that can propagate in vacuum --- then we must depart from the ECSK theory. 
So far two attempts have been proposed 
to construct gravitational theory with dynamical torsion. 
One is the quadratic Lagrangian approach 
to Poincar\'e gauge theory initiated by Hayashi [4] in 1968. 
Further Hayashi and Shirafuji [5-7], 
and Hehl and Von der Heyde [8] pursued this approach.
In this theory the torsion satisfies second-order differential equations 
and can propagate in vacuum. 
The other is the theory of new general relativity 
investigated from geometrical and observational point of view 
by Hayashi and Shirafuji [9]. 
In this theory the concept of absolute parallelism 
plays the fundamental role: It requires a vanishing local connection and 
hence a vanishing curvature, and attributes gravity to the torsion alone.
\par
In this paper we propose a new approach to dynamical torsion. 
We suppose that spacetime has a locally Lorentzian metric 
and a nonsymmetric connection, 
and introduce sixteen fields forming a quadruplet of basis vectors, 
which are not necessarily orthogonal with each other. 
Next we require a symmetric condition for the local connection 
with respect to this basis. 
Then we find that torsion can be expressed 
in terms of first-order derivatives of the sixteen fields: 
Accordingly, we call the sixteen fields ``torsion potential''.
The curvature is not vanishing in the present theory in contrast with 
new general relativity based on absolute parallelism. 
Starting from the Einstein-Cartan action with torsion, 
but regarding the metric tensor 
and the torsion potential as independent variables, 
we obtain the Einstein equation and the torsion field equation.
\par
In Sec.2 we show how to introduce the torsion potential 
and how to represent torsion and connection by it. 
In Sec.3 we derive field equations for the metric tensor 
and the torsion potential. 
In Sec.4 we discuss local linear transformations of the torsion potential, 
which differ from local Lorentz transformations. 
In Sec.5 we give a simple example in which 
the torsion potential is constructed from a scalar field. 
In the final section we summarize our results.
\par
%
%
\newsection{Torsion potential}
Let us suppose that spacetime has a locally Lorentzian metric 
$\mbox{\boldmath $g$}$ 
and a nonsymmetric connection $\mbox{\boldmath $\Gamma$}$. 
We write the connection coefficients ${\mit\Gamma}^\mu{}_{\rho\sigma}$ as
\be
{\mit\Gamma}^\mu{}_{\rho\sigma} 
= \Gammao{}^\mu{}_{\rho\sigma} + S^\mu{}_{\rho\sigma}\, ,
\label{eq2.1}
\ee
where $\Gammao{}^\mu{}_{\rho\sigma}$ 
is the torsionless connection coefficients symmetric in $\rho$ and $\sigma$, 
and $S^\mu{}_{\rho\sigma}$ is the contorsion tensor 
defined by the torsion tensor $T^\mu{}_{\rho\sigma}\,
(:=\!{\mit\Gamma}^\mu{}_{\rho\sigma}\!-\!{\mit\Gamma}^\mu{}_{\sigma\rho})$ 
in the form
\be
S_{\mu\rho\sigma} := {1 \over 2}\, 
\left( T_{\mu\rho\sigma} + T_{\rho\sigma\mu} + T_{\sigma\rho\mu} \right)\, ,
\label{eq2.2}
\ee
satisfying $S_{\mu\rho\sigma}\!=\!- S_{\rho\mu\sigma}$. 
When one takes variation 
of the Einstein-Cartan action with torsion, 
one usually regards the metric tensor $g_{\mu\nu}$ 
and the contorsion tensor $S_{\mu\nu\rho}$ 
(or equivalently, the torsion tensor $T_{\mu\nu\rho}$) 
as independent variables. 
This seems unnatural, however, because these two tensors are different 
in nature from each other: In fact, 
the metric tensor plays the role of a gravitational potential, 
while the contorsion tensor is treated as a quantity like a force. 
It is thus desirable to find a potential for torsion, 
namely a set of fields whose first-order derivatives define the torsion.
\par
For this purpose, let us introduce a linear transformation
\be
\ebase_A = t^\mu{}_A \, \dbase_\mu \, , \hspace{2cm}
\theta^A = t^A{}_\mu \, dx^\mu \, ,
\label{eq2.3}
\ee
where $\dbase_\mu$ is the natural basis and $\ebase_A$ an arbitrary one 
with $\theta^A$ being the 1-form dual to $\ebase_A \ (A \!=\! 0 \!\sim\! 3)$. 
Here $t^A{}_\mu$ and its inverse $t^\mu{}_A$ 
are some functions of coordinates 
satisfying the condition that the determinant 
$t \!=\!\det(t^A{}_\mu)$ never vanishes. Obviously they satisfy
\be
t^\mu{}_A \, t^A{}_\nu = \delta^\mu_\nu \, , \hspace{2cm}
t^A{}_\mu \, t^\mu{}_B = \delta^A_B \, .
\label{eq2.4}
\ee
On this new basis the metric tensor becomes
\be
g_{AB} = \ebase_A \cdot \ebase_B = t^\mu{}_A \, t^\nu{}_B \, g_{\mu\nu}\, .
\label{eq2.5}
\ee
Here we do {\it not assume} that $g_{AB}$ coincides with the Minkowski metric 
$\eta_{AB}\!=\!{\rm diag}\,(+1,-1,-1,-1)$. Accordingly, 
the field $t^A{}_\mu$ is not a tetrad in the ordinary sense.
\par
Now we define connection coefficients 
with respect to the bases $\dbase_\mu$ and $\ebase_A$ as follows:
\be
\nabla \, \dbase_\nu = {\mit\Gamma}^\mu{}_\nu \, \dbase_\mu \, , \hspace{2cm}
\nabla \, \ebase_B = \theta^A{}_B \, \ebase_A
\label{eq2.6}
\ee
with the connection 1-forms ${\mit\Gamma}^\mu{}_\nu$ and $\theta^A{}_B$ being
\ba
{\mit\Gamma}^\mu{}_\nu &\!\!\! := &\!\!\! 
{\mit\Gamma}^\mu{}_{\nu\lambda}\, dx^\lambda 
= {\mit\Gamma}^\mu{}_{\nu A}\, \theta^A \, ,
\label{eq2.7} \\[.5mm]
\theta^A{}_B &\!\!\! := &\!\!\! \theta^A{}_{B \lambda}\, dx^\lambda 
= \theta^A{}_{BC}\, \theta^C \, ,
\label{eq2.8}
\ea
respectively. The two connection coefficients are related to each other by
\be
\theta^A{}_{BC} = t^A{}_\mu \, t^\nu{}_B \, t^\lambda{}_C \, 
{\mit\Gamma}^\mu{}_{\nu\lambda} - t^\nu{}_B \, t^\lambda{}_C \, 
\partial_\lambda \, t^A{}_\nu \, .
\label{eq2.9}
\ee
In conformity with this, the total covariant derivative 
of $t^A{}_\mu$ with respect to indices $A$ and $\mu$ is vanishing:
\be
\left\{
\begin{array}{rcl}
{\cal D}_\lambda \, t^A{}_\nu &\!\!\! := &\!\!\! 
\partial_\lambda \, t^A{}_\nu - {\mit\Gamma}^\mu{}_{\nu\lambda}\, t^A{}_\mu 
+ \theta^A{}_{B \lambda}\, t^B{}_\nu = 0 \, , \\[2mm]
{\cal D}_\lambda \, t^\mu{}_B &\!\!\! := &\!\!\! 
\partial_\lambda \, t^\mu{}_B + {\mit\Gamma}^\mu{}_{\nu\lambda}\, t^\nu{}_B 
- \theta^A{}_{B \lambda}\, t^\mu{}_A = 0 \, .
\end{array}
\right.
\label{eq2.10}
\ee
\par
The commutator of the basis $\ebase_A$ 
and the exterior derivative of the 1-form $\theta^A$ are respectively given by
\be
\left[ \, \ebase_B \ , \ebase_C \, \right] = f^A{}_{BC}\, \ebase_A \, ,
\hspace{2cm}
d \theta^A = - {1 \over 2}\, f^A{}_{BC}\, \theta^B \! \wedge \theta^C \, ,
\label{eq2.11}
\ee
where $f^A{}_{BC}$, 
which are called the Ricci rotation coefficients in the usual tetrad case, 
are given by
\be
f^A{}_{BC} = \left( t^\mu{}_B \, t^\nu{}_C - t^\mu{}_C \, t^\nu{}_B \right) 
\partial_\nu \, t^A{}_\mu \, .
\label{eq2.12}
\ee
On the basis $\ebase_A$ the torsion 2-form $T^A$ 
and the curvature 2-form $R^A{}_B$ are defined by
\ba
T^A &\!\!\! := &\!\!\! d \theta^A + \theta^A{}_B \! \wedge \theta^B 
= - {1 \over 2}\, T^A{}_{BC}\, \theta^B \! \wedge \theta^C \, ,
\label{eq2.13} \\[1mm]
R^A{}_B &\!\!\! := &\!\!\! d \theta^A{}_B + \theta^A{}_C 
\wedge \theta^C{}_B = {1 \over 2}\, R^A{}_{BCD}\, 
\theta^C \! \wedge \theta^D \, ,
\label{eq2.14}
\ea
where the torsion tensor and the curvature tensor are respectively given by
\ba
T^A{}_{BC} &\!\!\! = &\!\!\! 
t^A{}_\mu \, t^\nu{}_B \, t^\lambda{}_C \, T^\mu{}_{\nu\lambda} 
= \theta^A{}_{BC} - \theta^A{}_{CB} + f^A{}_{BC}\, ,
\label{eq2.15} \\[1mm]
R^A{}_{BCD} &\!\!\! = &\!\!\! 
t^A{}_\mu \, t^\nu{}_B \, t^\rho{}_C \, t^\sigma{}_D \, 
R^\mu{}_{\nu\rho\sigma}\, .
\label{eq2.16}
\ea
We notice that since $g_{AB}\! \neq \!\eta_{AB}$ by assumption, 
$\theta_{ABC}\! \neq \! - \theta_{BAC}$, and therefore that 
$\theta_{ABC}$ is not a spin connection. 
In order to compare with the torsionless case 
and for convenience of later use, 
let us separate the curvature tensor $R^\mu{}_{\nu\rho\sigma}$ 
into torsionless and torsion parts,
\be
R^\mu{}_{\nu\rho\sigma} = \Ro{}^\mu{}_{\nu\rho\sigma} 
+ \RT{}^\mu{}_{\nu\rho\sigma}\, ,
\label{eq2.17}
\ee
where $\Ro{}^\mu{}_{\nu\rho\sigma}$ is the curvature tensor 
for the symmetric connection,
\be
\Ro{}^\mu{}_{\nu\rho\sigma} = \partial_\rho \Gammao{}^\mu{}_{\nu\sigma} 
- \partial_\sigma \Gammao{}^\mu{}_{\nu\rho} 
+ \Gammao{}^\mu{}_{\lambda\rho} \Gammao{}^\lambda{}_{\nu\sigma} 
- \Gammao{}^\mu{}_{\lambda\sigma} \Gammao{}^\lambda{}_{\nu\rho}
\label{eq2.18}
\ee
and $\RT{}^\mu{}_{\nu\rho\sigma}$ is the torsion part of the curvature tensor 
defined by
\be
\RT{}^\mu{}_{\nu\rho\sigma} := \nablao_\rho S^\mu{}_{\nu\sigma} 
- \nablao_\sigma S^\mu{}_{\nu\rho} 
+ S^\mu{}_{\lambda\rho} S^\lambda{}_{\nu\sigma} 
- S^\mu{}_{\lambda\sigma} S^\lambda{}_{\nu\rho}\, .
\label{eq2.19}
\ee
Here $\nablao_\rho$ denotes the covariant derivative with respect to 
$\Gammao{}^\mu{}_{\nu\rho}$.
\par
Until now we have not yet imposed any restriction on 
$t^A{}_\mu$ and $\theta^A{}_{BC}$. From now on, however, 
we suppose that the following two conditions are satisfied:
\par
\noindent
(i) \ \ The metric condition
\be
{\cal D}_\lambda g_{\mu\nu} = 0 \, .
\label{eq2.20}
\ee
\noindent
(ii) \ \ The symmetric condition
\be
\theta^A{}_{BC} = \theta^A{}_{CB} \hspace{1cm} {\rm or \ \, equivalently} 
\hspace{1cm} \theta^A{}_B \! \wedge \theta^B = 0 \, .
\label{eq2.21}
\ee
The condition (i) implies that $\Gammao{}^\mu{}_{\rho\sigma}$ 
is nothing but the Christoffel symbol. Using the differential operators 
$d_C := t^\mu{}_C \, \partial_\mu$ and 
$\ {\cal D}_C := t^\mu{}_C \, {\cal D}_\mu$, 
we see from (\ref{eq2.5}), (\ref{eq2.10}) and (\ref{eq2.20}) that
\be
{\cal D}_C g_{AB} = 0 \, .
\label{eq2.22}
\ee
The condition (ii) can be regarded as 
a generalization of absolute parallelism, 
because we have $\theta_{ABC}\!=\! 0$ when $g_{AB}\!= \!\eta_{AB}$. 
Following the procedure to obtain the Christoffel symbol 
$\Gammao{}^\mu{}_{\rho\sigma}$ from $g_{\mu\nu}$, 
we can express $\theta^A{}_{BC}$ as
\be
\theta^A{}_{BC} 
= {1 \over 2}\, g^{AD} \left( d_B g_{CD} + d_C g_{BD} - d_D g_{BC} \right)
\label{eq2.23}
\ee
by means of (\ref{eq2.21}) and (\ref{eq2.22}).
\par
Like (\ref{eq2.1}), $\theta^A{}_{BC}$ can be written as
\be
\theta^A{}_{BC} = \thetao{}^A{}_{BC} + S^A{}_{BC}\, ,
\label{eq2.24}
\ee
where $\thetao{}^A{}_{BC}$ are obtained 
by using $\Gammao{}^\mu{}_{\rho\sigma}$ 
instead of ${\mit\Gamma}^\mu{}_{\rho\sigma}$ in (\ref{eq2.9}). 
It should be noted that $\thetao{}^A{}_{BC}\! \neq \!\thetao{}^A{}_{CB}$ 
unless the torsion vanishes identically. 
$S^A{}_{BC}$ are the components of the contorsion tensor 
with respect to the basis $\ebase_A$, 
being related to $S^\mu{}_{\nu\lambda}$ by
\be
S^A{}_{BC} = t^A{}_\mu \, t^\nu{}_B \, t^\lambda{}_C \, 
S^\mu{}_{\nu\lambda}\, .
\label{eq2.25}
\ee
Applying (\ref{eq2.21}) to (\ref{eq2.13}) and (\ref{eq2.15}), we get
\be
T^A = d \theta^A \hspace{1cm} {\rm and} \hspace{1cm}
T^A{}_{BC} = f^A{}_{BC}\, .
\label{eq2.26}
\ee
According to (\ref{eq2.12}), the latter equation of (\ref{eq2.26}) 
means that the torsion tensor is expressed by first-order derivatives 
of $t^A{}_\mu$: For example, the torsion tensor $T^\mu{}_{\nu\lambda}$ 
is given by
\be
T^\mu{}_{\nu\lambda} = t^\mu{}_A \left( \partial_\lambda t^A{}_\nu 
- \partial_\nu t^A{}_\lambda \right)\, .
\label{eq2.27}
\ee
We call the set of sixteen functions $t^A{}_\mu$ ``the torsion potential'', 
since it is changed like a set of covariant vectors 
under coordinate transformations 
and its first-order derivatives define 
the torsion and hence the contorsion tensor.
\par
%
%
\newsection{Field equations}
In this section we give field equations 
for the metric tensor and the torsion potential. 
Let us start from the Lagrangian
\be
{\cal L}_G = - {1 \over 2 \kappa}\,\sqrt{-g}\, R 
= - {1 \over 2 \kappa}\,\sqrt{-g}\, \Bigl(\Ro + \RT \Bigr)\, ,
\label{eq3.1}
\ee
where $\Ro$ is the Riemannian scalar curvature 
made of the metric tensor $g_{\mu\nu}$, 
and $\RT$ is the torsion part of the scalar curvature which contains 
$g_{\mu\nu}$ and $t^A{}_\mu$. 
Here $\kappa = 8 \pi G$ is the Einstein gravitational constant. 
According to (\ref{eq2.19}), we have
\be
\RT = 2\,\nablao_\mu S^{\mu\rho}{}_\rho 
- S^{\mu\rho}{}_\rho \,S_\mu{}^\sigma{}_\sigma 
- S^\mu{}_{\rho\sigma} \,S^{\rho\sigma}{}_\mu \, .
\label{eq3.2}
\ee
Taking variation of (\ref{eq3.1}) with respect to $g_{\mu\nu}$, we get
\be
\Go{}^{\mu\nu} = S^{\mu\rho}{}_\rho S^{\nu\sigma}{}_\sigma 
- S^{\mu\rho\sigma} S^\nu{}_{\sigma\rho} 
- {1 \over 2}\, g^{\mu\nu} 
S^{\lambda\rho}{}_\rho \,S_\lambda{}^\sigma{}_\sigma 
- {1 \over 2}\, g^{\mu\nu} 
S^\lambda{}_{\rho\sigma} S^{\rho\sigma}{}_\lambda \, ,
\label{eq3.3}
\ee
where $\Go{}^{\mu\nu}$ is the torsion-free Einstein tensor. 
The right-hand side of (\ref{eq3.3}) is the 
energy-momentum tensor of the torsion potential $t^A{}_\mu$, 
which now plays the role of a matter field. 
If $S_{\mu\rho\sigma}$ vanishes, (\ref{eq3.3}) 
returns to the vacuum Einstein equation. 
The contraction of (\ref{eq3.3}) with $g_{\mu\nu}$ gives
\be
\Ro = S^\mu{}_{\rho\sigma} S^{\rho\sigma}{}_\mu 
+ S^{\mu\rho}{}_\rho \,S_\mu{}^\sigma{}_\sigma 
= 2 \, \nablao_\mu S^{\mu\nu}{}_\nu - \RT \, ,
\label{eq3.4}
\ee
where we have used (\ref{eq3.2}) in the second equation.
\par
Taking variation of (\ref{eq3.1}) with respect to 
the torsion potential $t^A{}_\mu$, we get another field equation
\ba
S^\mu{}_{\rho\sigma} S^{\rho\sigma}{}_\nu t^\nu{}_A &\!\!\! + &\!\!\! 
S^\mu{}_{\nu\rho} S^{\rho\sigma}{}_\sigma t^\nu{}_A 
- S^\mu{}_{\rho\nu} S^{\rho\sigma}{}_\sigma t^\nu{}_A 
- \nablao_\rho S^{\rho\mu}{}_\nu t^\nu{}_A 
+ S^{\mu\rho}{}_\nu \nablao_\rho t^\nu{}_A \nonu
{} &\!\!\! + &\!\!\! \nablao_\nu S^{\nu\rho}{}_\rho t^\mu{}_A 
+ S^{\nu\rho}{}_\rho \nablao_\nu t^\mu{}_A 
- S^{\mu\rho}{}_\rho \nablao_\nu t^\nu{}_A 
- \nablao_\nu S^{\mu\rho}{}_\rho t^\nu{}_A = 0 \, .
\label{eq3.5}
\ea
The contraction of (\ref{eq3.5}) with $t^A{}_\mu$ gives
\be
\nablao_\mu S^{\mu\nu}{}_\nu = 0 \, .
\label{eq3.6}
\ee
Applying (\ref{eq3.6}) to (\ref{eq3.4}), we have
\be
R = \Ro + \RT = 0 \, .
\label{eq3.7}
\ee
Further, with the help of (\ref{eq3.4}) Einstein equation (\ref{eq3.3}) 
can be rewritten as
\be
\Ro{}^{\mu\nu} = S^{\mu\rho}{}_\rho \,S^{\nu\sigma}{}_\sigma 
- S^{\mu\rho\sigma} S^\nu{}_{\sigma\rho}\, .
\label{eq3.8}
\ee
\par
We remark that the equation (\ref{eq3.5}) is a tensor equation 
under coordinate transformations, while it is not invariant 
under local linear transformations which we shall discuss in the next section. 
We rewrite (\ref{eq3.5}) and (\ref{eq3.6}) 
on the basis $\ebase_A$:
\ba
\Do_B S^{AB}{}_C &\!\!\! - &\!\!\! \Do_C S^{AB}{}_B 
+ \thetao{}^B{}_{CD} S^{AD}{}_B + \thetao{}^A{}_{BC} S^{BD}{}_D \nonu
{} &\!\!\! - &\!\!\! \thetao{}^A{}_{BD} S^{BD}{}_C 
- \thetao{}^B{}_{CB} S^{AD}{}_D = 0 \, ,
\label{eq3.9} \\[2mm]
\Do_B S^{BD}{}_D &\!\!\! = &\!\!\! 0 \, ,
\label{eq3.10}
\ea
where $\Do_B$ is the total covariant derivative operator 
using $\Gammao{}^\mu{}_{\nu B}$ and $\thetao{}^A{}_{CB}$. 
The equation (\ref{eq3.9}) is not a tensor equation 
because it contains connection $\thetao{}^A{}_{BC}$, 
while (\ref{eq3.10}) is really a tensor one 
since its connection terms are canceled out with each other 
when the indices $A$ and $C$ are contracted in (\ref{eq3.9}).
\par
%
%
\newsection{Local linear transformations of the torsion potential}
In this section we consider properties 
of the torsion potential under local linear transformations. 
Let $L^A{}_B$ be some transformation functions which define
\be
{\hat t}^A{}_\mu = L^A{}_B \, t^B{}_\mu \, , \hspace{2cm}
{\hat t}^\mu{}_B = L^{-1}{\,{}^A}_B \, t^\mu{}_A \, .
\label{eq4.1}
\ee
The 1-form $\theta^A$ and the connection $\theta^A{}_B$ 
are then transformed like
\ba
{\hat \theta}^A &\!\!\! = &\!\!\! L^A{}_B \, \theta^B \, ,
\label{eq4.2} \\[1mm]
{\hat \theta}^A{}_B &\!\!\! = &\!\!\! 
L^A{}_C \, L^{-1}{\,{}^D}_B \, \theta^C{}_D 
+ L^A{}_C \, d L^{-1}{\,{}^C}_B \, .
\label{eq4.3}
\ea
Here ${\hat \theta}^A{}_B \,(= {\hat \theta}^A{}_{BC} {\hat \theta}^C)$ 
is a new connection, which we require to satisfy the condition (ii); namely,
\be
{\hat \theta}^A{}_{BC} = {\hat \theta}^A{}_{CB} \hspace{1cm}
{\rm or \ \, equivalently}
\hspace{1cm} {\hat \theta}^A{}_B \! \wedge {\hat \theta}^B = 0 \, .
\label{eq4.4}
\ee
Then we see that the transformation functions 
$L^A{}_B$ must satisfy the constraint
\be
d_B L^A{}_C = d_C L^A{}_B \hspace{1cm} {\rm or} \hspace{1cm} 
L^B{}_C d_D L^{-1}{\,{}^A}_B = L^B{}_D d_C L^{-1}{\,{}^A}_B
\label{eq4.5}
\ee
with $d_B = t^\mu{}_B \, \partial_\mu$. 
Incidentally we remark that general coordinate transformations 
$J^\mu{}_\nu \!=\!(\partial x'^\mu / \partial x^\nu)$ 
automatically satisfy the similar condition 
$\partial_\rho J^\mu{}_\nu \!=\!\partial_\nu J^\mu{}_\rho$. 
Under transformations (\ref{eq4.1}) satisfying (\ref{eq4.5}), 
the torsion and the curvature 2-forms transform as follows:
\ba
{\hat T}^A &\!\!\! = &\!\!\! L^A{}_B \, T^B \, ,
\label{eq4.6} \\[1mm]
{\hat R}^A{}_B &\!\!\! = &\!\!\! L^A{}_C \, L^{-1}{\,{}^D}_B \, R^C{}_D \, .
\label{eq4.7}
\ea
\par
Next, we consider properties of the torsion equation (\ref{eq3.9}) 
under local linear transformations: The left-hand side of (\ref{eq3.9}) 
is changed like
\ba
& &\!\!\!\!\!\!\!{\widehat {\Do}}_B {\hat S}{}^{AB}{}_C 
- {\widehat {\Do}}_C {\hat S}{}^{AB}{}_B 
+ {\hat {\thetao}}{}^B{}_{CD} {\hat S}{}^{AD}{}_B 
+ {\hat {\thetao}}{}^A{}_{BC} {\hat S}{}^{BD}{}_D 
- {\hat {\thetao}}{}^A{}_{BD} {\hat S}{}^{BD}{}_C 
- {\hat {\thetao}}{}^B{}_{CB} {\hat S}{}^{AD}{}_D \nonu
= & &\!\!\!\!\!\!\! L^A{}_B L^{-1}{\,{}^D}_C \Bigl( 
\Do_M S^{BM}{}_D - \Do_D S^{BM}{}_M 
+ \thetao{}^M{}_{DN} S^{BN}{}_M + \thetao{}^B{}_{MD} S^{MN}{}_N \nonu
& & \hspace{1.8cm} - \, \thetao{}^B{}_{MN} S^{MN}{}_D 
- \thetao{}^M{}_{DM} S^{BN}{}_N \Bigr) \nonu
& &\!\!\!\!\!\!\! + \, d_D \left( L^A{}_B L^{-1}{\,{}^M}_C \right) S^{BD}{}_M 
- d_M \left( L^A{}_B L^{-1}{\,{}^M}_C \right) S^{BD}{}_D \, .
\label{eq4.8}
\ea
Thus, the torsion equation is invariant under (\ref{eq4.1}), 
if and only if $L^A{}_B$ satisfy the constraint,
\be
L^A{}_B d_D L^{-1}{\,{}^M}_C = L^A{}_D d_B L^{-1}{\,{}^M}_C \, .
\label{eq4.9}
\ee
We notice that contracting $A$ and $C$ in (\ref{eq4.9}) gives (\ref{eq4.5}).
\par
In order to see the consequence of (\ref{eq4.9}), 
let us consider infinitesimal transformations
\be
L^A{}_B = \delta^A_B + \varepsilon^A{}_B
\label{eq4.10}
\ee
with $|\varepsilon^A{}_B| \ll 1$. 
If (\ref{eq4.10}) satisfy the constraint (\ref{eq4.9}), 
we have
\be
d_B \, \varepsilon^A{}_C = 0 \, ,
\label{eq4.11}
\ee
which implies that $\varepsilon^A{}_B$ must be constant. 
Therefore we see that the invariance linear group of (\ref{eq3.9}) 
is the global $GL(4,R)$. 
This situation is to be compared with that in new general relativity 
where the invariance linear group is the global Lorentz group.
\par
%
%
\newsection{Scalar model of the torsion potential}
Now we consider as a simple example 
the torsion potential made of a scalar field. 
Namely, let us assume that the torsion potential takes the form
\be
t^A{}_\mu = \delta^A_\mu \varphi \, , \hspace{2cm}
t^\mu{}_A = \delta^\mu_A \varphi^{-1}
\label{eq5.1}
\ee
with $\varphi$ being a nonvanishing scalar field. 
Due to Kronecker's $\delta^A_\mu$ in (\ref{eq5.1}), 
the index $A$ is assumed to undergo the transformation generated by 
$L^A{}_B = \delta^A_\nu \delta^\mu_B \,(\partial x^\nu / \partial x'^\mu)$ 
whenever general coordinate transformations, 
$x^\mu \!\rightarrow \! x'^\mu$, are made. 
The metric tensor $g_{AB}$ is given by
\be
g_{AB} = \varphi^{-2}\, \delta^\mu_A \, \delta^\nu_B \, g_{\mu\nu}\, .
\label{eq5.2}
\ee
The torsion and the contorsion tensors are expressed by
\ba
T_{\mu\nu\lambda} &\!\!\! = &\!\!\! 
g_{\mu\nu} \varphi^{-1} \partial_\lambda \varphi 
- g_{\mu\lambda} \varphi^{-1} \partial_\nu \varphi \, ,
\label{eq5.3} \\[1mm]
S^{\mu\nu}{}_\lambda &\!\!\! = &\!\!\! 
\left( g^{\mu\rho} \delta^\nu_\lambda 
- g^{\nu\rho} \delta^\mu_\lambda \right) 
\varphi^{-1} \partial_\rho \varphi \, .
\label{eq5.4}
\ea
If we substitute $S^{\mu\nu}{}_\lambda$ of (\ref{eq5.4}) 
to the full equation (\ref{eq3.5}), we will encounter a difficulty, 
because $16$ equations cannot be satisfied 
by only one unknown function $\varphi$. 
Thus, we will choose a different way: 
Namely, we directly substitute $S^{\mu\nu}{}_\lambda$ 
to the Lagrangian (\ref{eq3.1}). Then we have
\be
{\cal L}_G = - {1 \over 2 \kappa}\,\sqrt{-g}\, \Bigl(\Ro - 6 g^{\mu\nu} 
\varphi^{-2} \partial_\mu \varphi \partial_\nu \varphi \Bigr)\, .
\label{eq5.5}
\ee
This is the Lagrangian for gravitational field 
coupled to a massless scalar field: 
In fact, if we put $\psi = \ln \varphi$, the second term of (\ref{eq5.5}) 
is just the kinetic term of the field $\psi$.
\par
Taking variation of (\ref{eq5.5}) with respect to $g_{\mu\nu}$ and $\varphi$, 
we get
\ba
& & \Go{}^{\mu\nu} = {6 \over \varphi^2}\, \left( 
g^{\mu\rho} g^{\nu\sigma} - {1 \over 2}\, g^{\mu\nu} g^{\rho\sigma} \right)
\partial_\rho \varphi \, \partial_\sigma \varphi \, ,
\label{eq5.6} \\[1mm]
& & \Boxo \, \varphi - {1 \over \varphi}\, 
g^{\mu\nu} \partial_\mu \varphi \, \partial_\nu \varphi = 0 \, ,
\label{eq5.7}
\ea
where $\Boxo$ denotes the torsionless d'Alembertian operator defined by 
$\Boxo = g^{\mu\nu} \nablao_\mu \nablao_\nu$. 
{}From (\ref{eq5.6}) and (\ref{eq5.7}) it follows
\ba
& & \Ro = {6 \over \varphi^2}\, g^{\mu\nu} 
\partial_\mu \varphi \, \partial_\nu \varphi \, ,
\label{eq5.8} \\[1mm]
& & \Boxo \, \varphi - {1 \over 6}\, \Ro \, \varphi = 0 \, .
\label{eq5.9}
\ea
If $\Ro = 0$, we get the massless scalar field equation
\be
\Boxo \, \varphi = 0 \, .
\label{eq5.10}
\ee
Incidentally we remark that substituting (\ref{eq5.4}) to 
the contracted form (\ref{eq3.6}) gives (\ref{eq5.9}). 
\par
The form of (\ref{eq5.2}) recalls us to conformal rescaling. 
Define a new ``unphysical'' metric ${\hat g}_{\mu\nu}$ 
from the original ``physical'' metric by
\be
{\hat g}_{\mu\nu} = \varphi^{-2} g_{\mu\nu}\, ,
\label{eq5.11}
\ee
then the Riemannian scalar curvature 
made of the new metric ${\hat g}_{\mu\nu}$ is given by
\be
{\widehat {\Ro}} = \varphi^2 \Bigl( 
\Ro - 6 g^{\mu\nu} \varphi^{-2} \partial_\mu \varphi \partial_\nu \varphi 
+ 6 \Boxo \ln \varphi \Bigr)\, .
\label{eq5.12}
\ee
Owing to this, the Lagrangian (\ref{eq5.5}) can be rewritten as
\be
{\cal L}_G = - {1 \over 2 \kappa}\, \varphi^2 
\sqrt{- {\hat g}}\, {\widehat {\Ro}}\, ,
\label{eq5.13}
\ee
where a total divergence term is omitted. 
We see that the kinetic term of (\ref{eq5.5}) is now absorbed into 
the scalar curvature ${\widehat {\Ro}}$ made of ${\hat g}_{\mu\nu}$, 
and the Lagrangian has the form of dilaton gravity 
without the derivative terms of $\varphi$.
\par
%
%
\newsection{Summary}
We have constructed a geometrical framework of the gravitational theory 
with torsion by introducing the torsion potential $t^A{}_\mu$. 
The torsion acquires a dynamical property, 
being expressed by first-order derivatives of the torsion potential. 
Starting from the Einstein-Cartan action with torsion, 
we have obtained field equations for the metric tensor 
and the torsion potential. 
The full equation for the torsion potential 
is not invariant under local linear transformations of the torsion potential. 
However, its contracted form is found to be invariant. 
We have shown that the invariance group of the full equation 
is the global general linear group. 
In this paper we have proposed the framework of the torsion potential 
only in source-free case, and given a simple example 
where the torsion potential is described by a scalar field. 
As a next step we shall study the possibility that the torsion potential 
is constructed from a vector field or a spinor field.
\par
On the other hand, matter fields other than the torsion potential 
can be introduced into this framework as follows. 
Scalar and vector fields can be treated 
in the same manner as in general relativity. 
Spinor fields can be introduced with the help of the tetrad fields, 
which are different from the torsion potential $t^A{}_\mu$, 
assuming that the covariant differentiation for them 
is defined by the spin connection with torsion. 
Detailed consideration of matter couplings will be given elsewhere.
\par
%
%
\vspace{1cm}
\begin{center}
{\large {\bf Acknowledgments}}
\end{center}
\par
We would like to thank members of the theory group 
of Physics Department at Saitama University, 
especially Professor K. Mori, Professor Y. Tanii, Dr.\ S. Yamaguchi 
and Dr.\ T. Kim for encouragements.
\par
%
%

%
\end{document}